\documentclass[aps,twocolumn,superscriptaddress,showpacs]{revtex4}

\usepackage{epsf,latexsym}
\usepackage{times}

\newcommand{\ket}[1]{| #1 \rangle}

\newcommand{\ra}{{\rightarrow}}
\newcommand{\up}{{\uparrow}}
\newcommand{\down}{{\downarrow}}
\newcommand{\ignore}[1]{}

\newcommand{\be}{\begin{equation}}
\newcommand{\ee}{\end{equation}}

\def\CC{{\rm\kern.24em \vrule width.04em height1.46ex depth-.07ex
    \kern-.30em C}}
\def\P{{\rm I\kern-.25em P}}

\def\bbbone{{\mathchoice {\rm 1\mskip-4mu l} {\rm 1\mskip-4mu l}
{\rm 1\mskip-4.5mu l} {\rm 1\mskip-5mu l}}}

\def\bbbc{{\mathchoice {\setbox0=\hbox{$\displaystyle\rm C$}\hbox{\hbox
to0pt{\kern0.4\wd0\vrule height0.9\ht0\hss}\box0}}
{\setbox0=\hbox{$\textstyle\rm C$}\hbox{\hbox
to0pt{\kern0.4\wd0\vrule height0.9\ht0\hss}\box0}}
{\setbox0=\hbox{$\scriptstyle\rm C$}\hbox{\hbox
to0pt{\kern0.4\wd0\vrule height0.9\ht0\hss}\box0}}
{\setbox0=\hbox{$\scriptscriptstyle\rm C$}\hbox{\hbox
to0pt{\kern0.4\wd0\vrule height0.9\ht0\hss}\box0}}}}

\def\bbbz{{\mathchoice {\hbox{$\sf\textstyle Z\kern-0.4em Z$}}
{\hbox{$\sf\textstyle Z\kern-0.4em Z$}}
{\hbox{$\sf\scriptstyle Z\kern-0.3em Z$}}
{\hbox{$\sf\scriptscriptstyle Z\kern-0.2em Z$}}}}

\newcommand{\putfig}[2]{$$\leavevmode\hbox{\epsfxsize=#2 cm
   \epsffile{#1.eps}}$$}

\begin{document}
\title{Single spin measurement using spin-orbital entanglement}
\author{Radu Ionicioiu}
\affiliation{Institute for Scientific Interchange (ISI), Villa Gualino, Viale Settimio Severo 65, I-10133 Torino, Italy}
\author{A.E.~Popescu}
\affiliation{Department of Engineering, University of Cambridge, Cambridge CB2 1PZ, UK}

\begin{abstract}
Single spin measurement represents a major challenge for spin-based quantum computation. In this article we propose a new method for measuring the spin of a single electron confined in a quantum dot (QD). Our strategy is based on entangling (using unitary gates) the spin and orbital degrees of freedom. An {\em orbital qubit}, defined by a second, empty QD, is used as an ancilla and is prepared in a known initial state. Measuring the orbital qubit will reveal the state of the (unknown) initial spin qubit, hence reducing the problem to the easier task of single charge measurement. Since spin-charge conversion is done with unit probability, single-shot measurement of an electronic spin can be, in principle, achieved. We evaluate the robustness of our method against various sources of error and discuss possible implementations.
\end{abstract}

\maketitle

\section{Introduction}

A notoriously difficult task in quantum information processing (QIP) and spintronics is the measurement of a single electron spin. The spin of an excess electron in a quantum dot is a natural candidate for the implementation of a solid-state qubit, since its Hilbert space is inherently two-dimensional (it is generally assumed that the electron is in the ground state and that transitions to higher excited states are negligible). While implementations of state preparation and quantum gates for a spin qubit are in principle feasible \cite{loss_ddv}, the measurement of a single spin still represents a major challenge \cite{hollenberg}. Several ideas for spin measurement have been proposed \cite{bandy}, including scanning tunneling microscopy \cite{STM} and magnetic resonance force microscopy (MRFM) \cite{MRF}. Optical detection of a single spin has been experimentally demonstrated \cite{optical_detect1,optical_detect2}.

Very recently several groups have experimentally demonstrated single-spin measurement using various techniques and in different environments. These experiments include the detection of an individual electronic spin using magnetic resonance force microscopy with 25 nm spatial resolution \cite{rugar} and the ESR detection of a single electron spin in a silicon transistor \cite{xiao}. Closer to the setup proposed here, single-shot spin readout in a quantum dot has been achieved by Elzerman {\em et al.} \cite{elzerman}. Especially encouraging in this last experiment is the very long single-spin relaxation time of up to 0.85 ms in a 8 T magnetic field.

In this article we discuss a method for measuring the spin of a single electron confined to a QD which can be adapted, in principle, to other spin-qubit proposals. Our algorithm is based on entangling the spin qubit with an orbital qubit used as an ancilla and prepared in a known initial state. Measuring the orbital qubit will reveal the state of the (unknown) initial spin qubit. Thus, by mapping (internal) spin degrees of freedom into (external) orbital degrees of freedom (or modes), we reduce the problem of single spin measurement to that of detecting the location of a particle in a double QD system. For an electron, this later problem becomes equivalent to single charge measurement. In contrast to single spin detection which is challenging, especially in a solid state environment, single charge measurement is easier and has been experimentally demonstrated. A radio-frequency single electron transistor (SET) has been used to observe {\em real-time} single electron tunneling in a QD \cite{real_time_SET}, whereas in Ref.~\cite{2SET_detector} two cross-correlated SETs were used to detect the charge state of a double QD. Other methods for single spin measurement based on spin-charge conversion have been discussed in Refs.~\cite{loss_ddv,Pazy,kane_single_spin,martin,recher,friesen,engel}.

One of the earliest proposals for single spin measurement employing spin-to-charge conversion \cite{loss_ddv} uses an auxiliary quantum dot with a single electron prepared in a known spin state. Compared to the Loss and DiVincenzo proposal \cite{loss_ddv}, our scheme does not require the presence of a second electron in the ancilla quantum dot, thus avoiding the experimentally challenging preparation of a single spin in a known state. This relaxes the technological constraints and reduces the possible sources of error.

\section{General setup}

Suppose we want to measure the spin state of a single electron confined in a QD. A second, empty QD is located in its vicinity. We denote the two dots by 0 and 1 respectively, and we will also refer to them as {\em modes}. We assume that dot 1 is decoupled from dot 0 (i.e., there is no tunneling between the two dots) during the whole quantum computation process; dot 1 is used only to detect the final spin state of the electron located initially in dot 0. In QIP terminology, we can say that dot 1 is used as an {\em ancilla} for measuring the spin qubit in dot 0.

A spin-1/2 particle in two QDs can encode a {\em spin qubit} and an {\em orbital qubit}. The basis states of the orbital qubit (also known as {\em dual rail} qubit) are defined by wave-functions localized in the 0, and respectively 1, quantum dot. We denote the total particle state by $\ket{\sigma; k}\equiv \ket{\sigma}\otimes \ket{k}$, where $\sigma= \up, \down$ represents the spin and $k=0,1$ the modes; the full Hilbert space is ${\cal H}= {\rm span}\left\{ \ket{\sigma; k}\right\}$.

\begin{figure}
\putfig{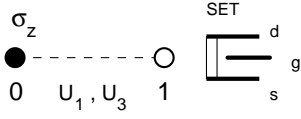}{4}
\caption{Measuring a single spin with a SET using spin-to-orbital conversion. A spin is initially located in site 0 (black dot). A second, empty dot (mode 1) resides in its vicinity, (open circle). After applying the unitary transformations $U_1U_2U_1$, a spin up (down) will be found with unit probability in dot 1 (0). Measuring the location of the electron with a SET is thus equivalent to measuring the initial spin state. An alternative design (experimentally implemented in \cite{2SET_detector}) uses two cross-correlated SETs (one for each QD) in order to avoid spurious detections due to background charges.}
\label{spin_set}
\end{figure}

A single electron tunneling between two dots (modes) is described by the hopping (tunneling) Hamiltonian
\be
H(t)= \tau(t)(a_{\sigma,0}^\dag a_{\sigma,1}+ \mbox{h.c.})
\ee
where $a_{\sigma,k}^\dag$ is the operator creating a particle with spin $\sigma$ in mode $k$. Since $H$ acts only on the mode degrees of freedom (tunneling does not change the spin), its action will induce Rabi oscillations between the two dots
\be
\ket{\sigma; 0}\ra \cos\theta\, \ket{\sigma; 0} + i\sin\theta\, \ket{\sigma; 1}
\ee
with $\theta= -\int \tau(t) dt/\hbar$. In the Hilbert space $\cal H$ defined above this is equivalent to a rotation around the $x$-axis in the $k$-subspace, $\bbbone_\sigma \otimes{\sf R_x}(\theta)_k$, where ${\sf R_x}(\theta)\equiv e^{i\theta \sigma_x}= \cos \theta\, \bbbone + i \sin \theta\, \sigma_x$.

Our proposed single spin measurement method involves four steps:\\
1.~couple the two QDs and allow the particle (situated initially in dot 0) to tunnel until is in an equal superposition between the two modes. This corresponds to a quarter of a Rabi oscillation between the dots, $\theta=\pi/4$, giving the transformation $U_1= \bbbone_\sigma \otimes {\sf R_x}(\pi/4)_k= {\rm diag}({\sf R_x}(\pi/4), {\sf R_x}(\pi/4))$;\\
2.~apply locally, on dot 0 only, a spin sign-flip $\sigma_z$. This leaves invariant the states $\ket{\up; 0}$, $\ket{\up; 1}$ and $\ket{\down; 1}$ and flips only the sign of $\ket{\down; 0}\ra -\ket{\down; 0}$. Hence the transformation is given by $U_2= {\rm diag}(1, 1, -1, 1)$. Since we assumed that dot 0 defines a spin qubit, this is a standard single-qubit operation and should be already available in any spin-based QC implementation;\\
3.~couple the two dots again as in step 1, $U_3= U_1$;\\
4.~detect (with a SET or otherwise) in which mode (QD) is the particle.

The succession of steps 1-3 implements in $\cal H$ a unitary transformation $U\equiv U_1 U_2 U_1= {\rm diag}(i\sigma_x, -\sigma_z)$. For a spin located initially in dot 0, $U$ induces the following mapping:
\begin{eqnarray}
\ket{\up; 0} \ra\ \ \ \ i\,\ket{\up; 1} \\
\ket{\down; 0} \ra\ -\ket{\down; 0}
\end{eqnarray}
This shows that a spin up (down) will always end up in dot 1 (0) with unit probability. The final step is to measure the electron location with a SET (see Fig.~\ref{spin_set}) \cite{set}. Hence, a {\em single-shot spin measurement} can be performed by detecting the charge of only one dot, e.g., QD 1 (in order to minimise interference between the SET and the spin qubit in dot 0).

From the above discussion we can see that our architecture is conceptually equivalent to a Mach-Zehnder interferometer (MZI). In a MZI a particle is coherently split by a beam-splitter, then propagates along two different paths and recombines again at a second beam-splitter (single particle interference). If the phase shift between the two branches is $\phi=0$ ($\pi$), an incoming particle in mode 0 will always (i.e., with unit probability) exit the interferometer in mode 1 (0). In our case a spin initially situated in QD 0 is coherently split between the two modes by $U_1$, then a spin phase shift is applied only to mode 0 by $U_2$, and then the two branches are recombined again by $U_3$. In the end, a spin-up (-down) particle will be recovered with unit probability in QD 1 (0) and the spin state can be measured by detecting the particle location with a SET coupled to QD 1. A similar scheme was used for mobile spins in a spintronic context for a mesoscopic spin Stern-Gerlach device \cite{pbs}.

\section{Error analysis}

In the previous description of our measurement method, all unitary gates and SETs were assumed to be ideal. Assuming that charge detection can be performed with high accuracy, then readout errors will be due to imperfect orbital and spin rotations. In this section we investigate how various gate errors affect the measurement accuracy.

We assume that due to imprecision in gate control the rotation angle in $U_1$ and $U_3$ will be different from the ideal value of $\pi/4$. The transformations induced by imperfect gates (marked with $'$ in the following) will be $U_1'= \bbbone_\sigma \otimes {\sf R_x}(\theta_1)_k$ and $U_3'= \bbbone_\sigma \otimes {\sf R_x}(\theta_2)_k$. The analysis of $U_2$ is more subtle. A $\sigma_z$ term in the Hamiltonian acting on spin is equivalent (up to a general $\pi/2$ phase) to a ${\sf R_z}(\pi)$ rotation, where ${\sf R_z}(\phi)\equiv e^{i\phi \sigma_z}$. However, in the 4-dimensional Hilbert space $\cal H$ of both orbital and spin degrees of freedom, this general phase appears only on mode 0, hence it is equivalent to a conditional gate between mode and spin. A general (i.e., non-ideal) expression for $U_2$ will include both the rotation angle $\phi$ and the extra phase $\psi$ induced by a spin sign-flip between orbital and spin degrees of freedom: $U_2'={\rm diag}(e^{i(\psi-\phi/2)}, 1, e^{i(\psi+\phi/2)}, 1)$. The ideal case corresponds to $\phi=\pi$ and $\psi=\pi/2$.
\begin{figure}
\putfig{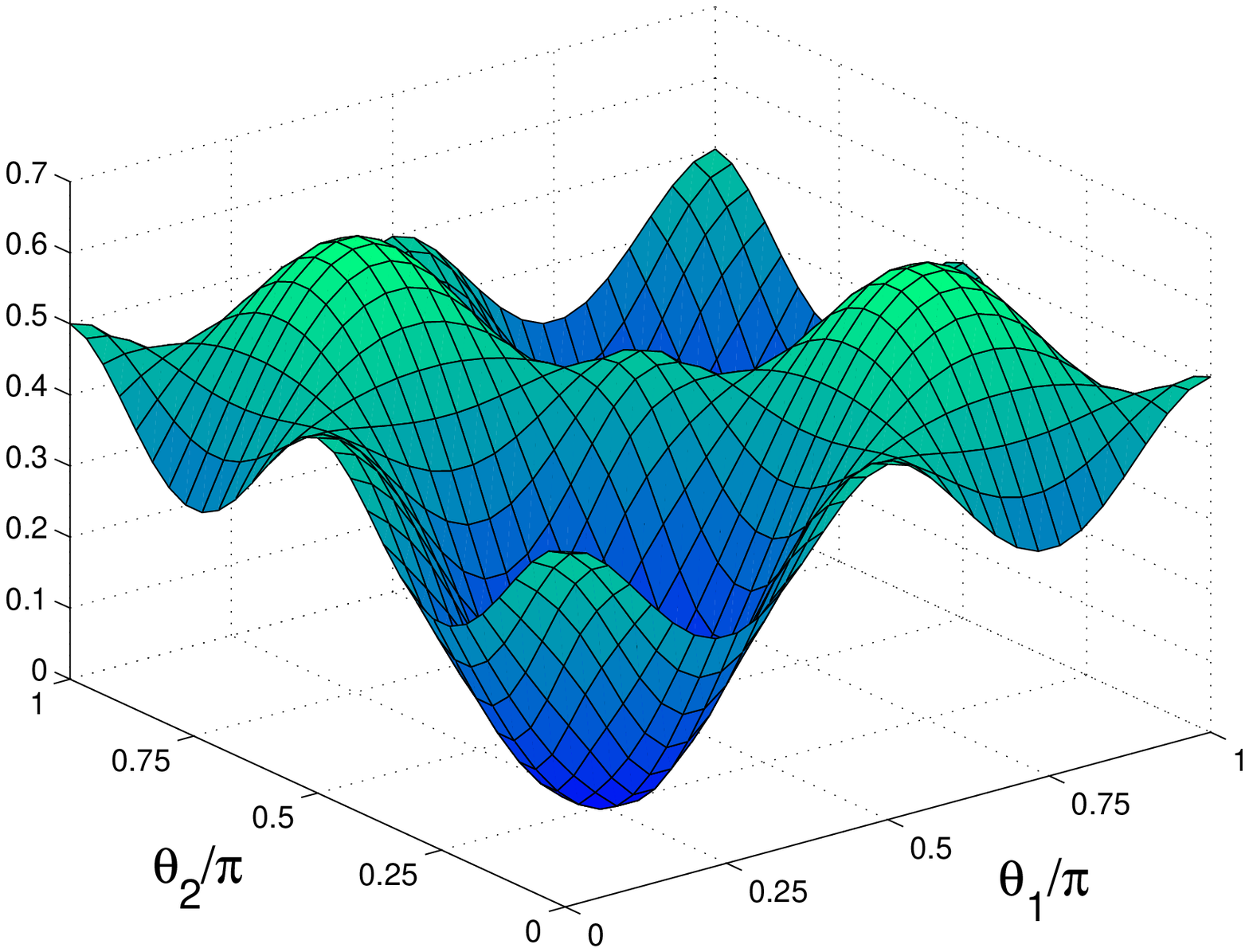}{7}
\putfig{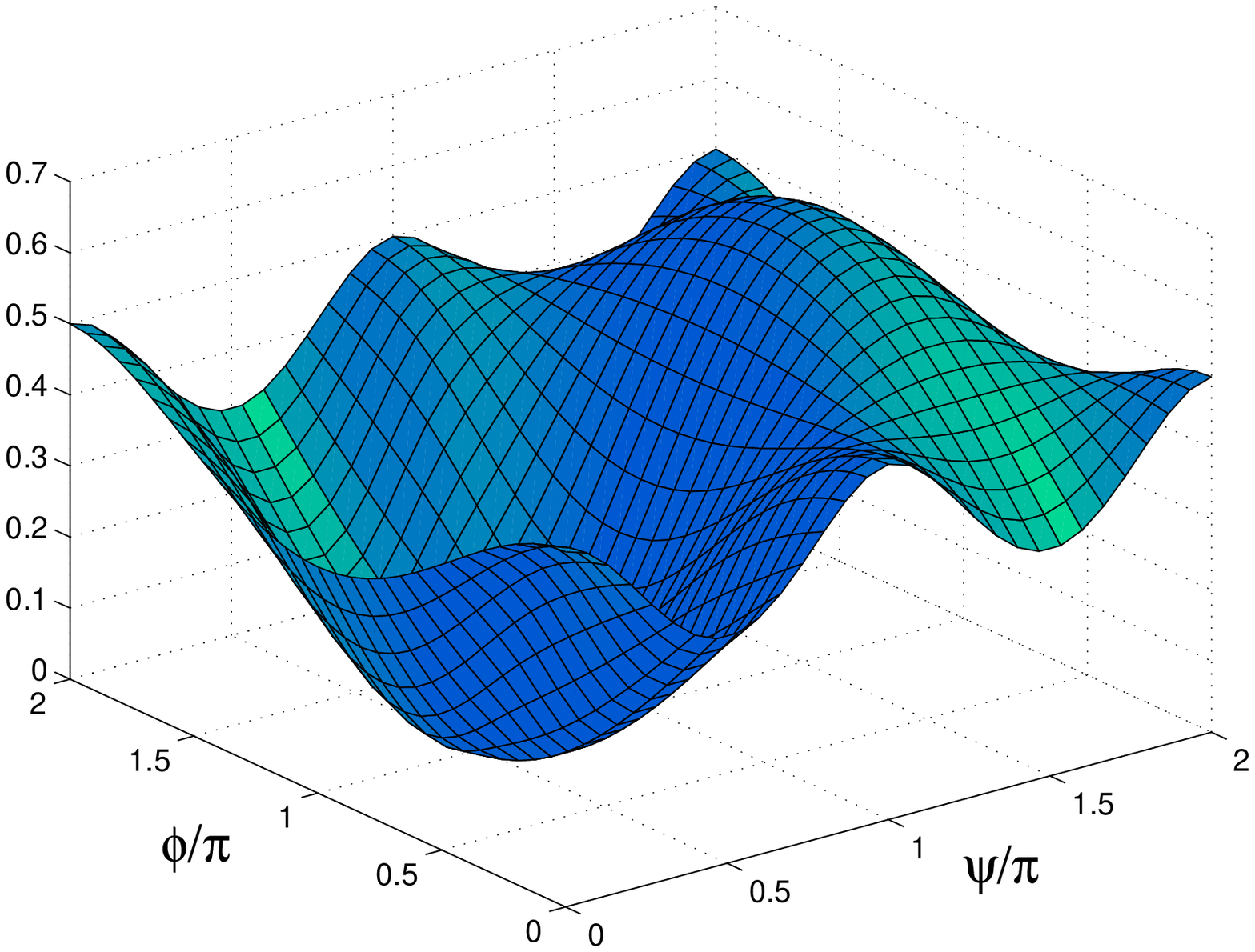}{7}
\putfig{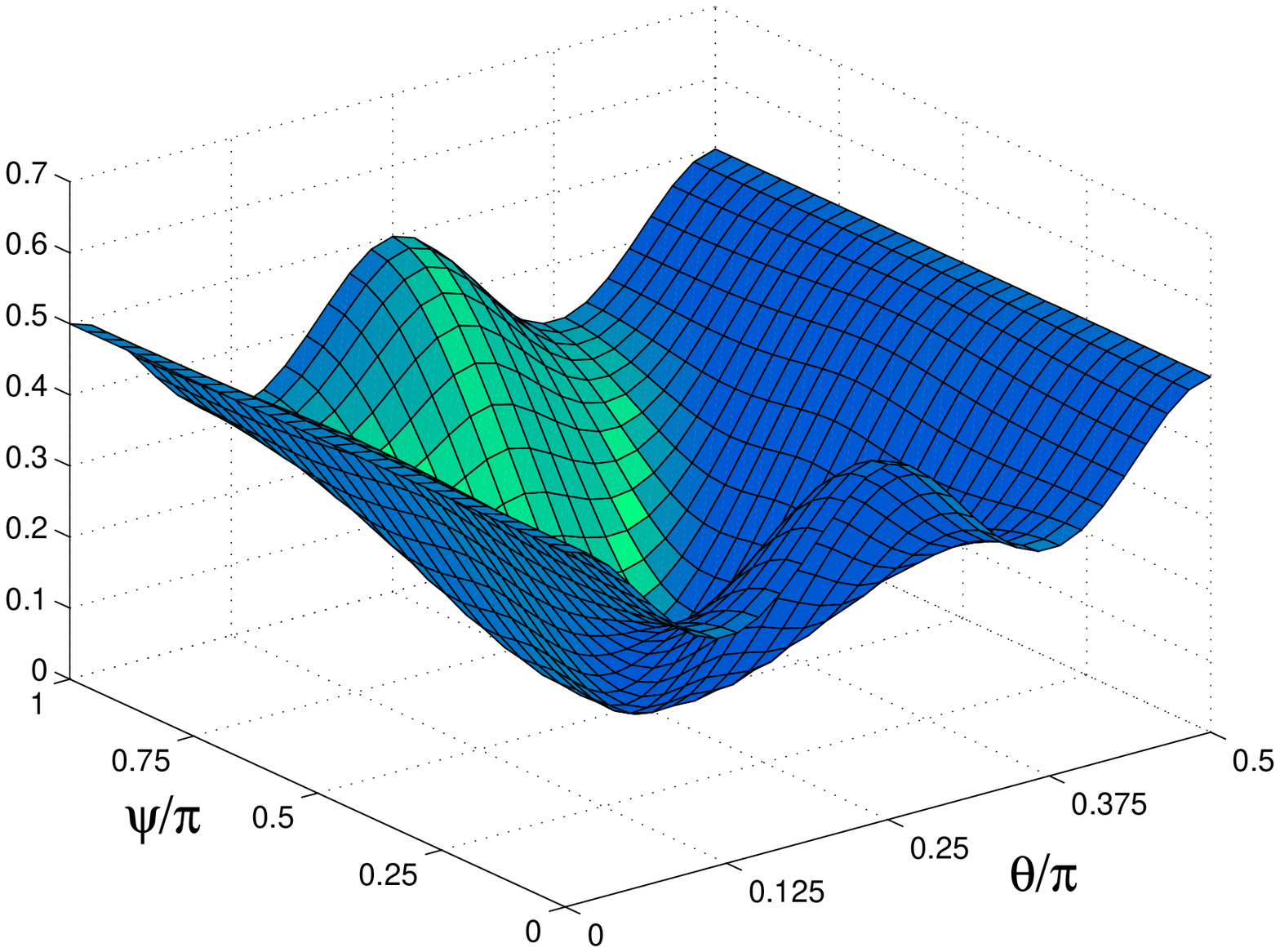}{7}
\caption{The average absolute error $\bar{\cal E}$ as a function of gate parameters: (a) $\bar{\cal E}(\theta_1, \theta_2)$ for $\psi=\pi/2$, $\phi=\pi$; (b) $\bar{\cal E}(\psi, \phi)$ with $\theta_1=\theta_2=\pi/4$; (c) $\bar{\cal E}(\theta, \psi)$ for $\theta_1=\theta_2= \theta$ and $\phi= 2\psi$.}
\label{errors}
\end{figure}

Suppose we want to measure an arbitrary spin state $\ket{\Psi_{in}}= (\cos\frac\delta 2 \ket{\up}+ e^{i\gamma}\sin\frac \delta 2 \ket{\down})\otimes\ket{0}$ (a spin superposition in dot 0). An ideal measuring apparatus would give the following probabilities:
\be
p_\up^{ideal}= \cos^2(\delta/2) \ \ \ \ , \ \ \ \ p_\down^{ideal}= \sin^2(\delta/2)
\ee

After applying a non-ideal sequence of gates $U'=U_3'U_2'U_1'$, the initial state $\ket{\Psi_{in}}$ is mapped into the output state:
\be
\ket{\Psi_{out}}= (f_1\ket{\up}+ f_2\ket{\down})\otimes \ket{0}+ (g_1\ket{\up}+ g_2\ket{\down})\otimes \ket{1}
\ee
where $f_i$ and $g_i$ are functions of $\theta_i$, $\phi$, $\psi$ and $\delta$ and can be read directly from the matrix elements of $U'$. Contrary to the ideal case, there will be a non-zero probability that a spin up (down) electron will end up in dot 0 (1).

Following the readout algorithm described above, one concludes that the probability of an electron ending up in dot 0 (1) originates from spin-down (up) part of the initial spin superposition. Therefore, performing an imperfect measurement (described by $U'$) will assign the following probabilities to the two basis spin states:
\begin{eqnarray}
p_\up= |g_1|^2+ |g_2|^2= \sin^2(\theta_1-\theta_2)+ A \\
p_\down= |f_1|^2+ |f_2|^2= \cos^2(\theta_1-\theta_2)- A
\end{eqnarray}
where
\[
A= \frac{1}{2} \sin 2\theta_1 \sin 2\theta_2 \left( 1+ \cos\psi \cos\frac{\phi}{2}+ \sin\psi \sin\frac{\phi}{2} \cos \delta \right)
\]
Note that both $p_\up$ and $p_\down$ are independent of the relative phase $e^{i\gamma}$ between the spin-up and spin down component of the initial state $\ket{\Psi_{in}}$. We define the {\em measurement error} as ${\cal E}= p_\up- p_\up^{ideal}= p_\down^{ideal}- p_\down$. For a fixed set of gate parameters $\Theta\equiv (\theta_1, \theta_2, \psi, \phi)$, the minimum (maximum) error over all input states occurs for $\delta=0$ ($\delta=\pi$, respectively), corresponding to the basis state $\ket{\up}$ ($\ket{\down}$). A physically significant function is the average (over all possible input states) of the absolute error $\bar{\cal E}(\Theta)= \frac 1 \pi \int_0^\pi |{\cal E}(\Theta,\delta)| d\delta$. In Figure \ref{errors} we plot $\bar{\cal E}$ as a function of various gate parameters.

\section{Implementations}

Experimentally, our method for spin-charge conversion is based on two requirements: (i) coherent control of particle tunneling between two quantum dots (steps 1 and 3); and (ii) fast switching of local fields required to enact the $\sigma_z$ spin-flip in dot 0 (step 2).

Coherent manipulation of charge tunneling in a double quantum dot has been experimentally demonstrated on a sub-nanosecond time scale \cite{doubleQD}. As described in Ref.~\cite{doubleQD}, the control is realized with rectangular voltage pulses applied to the drain electrode. The transformation $U_1$ required in steps 1 and 3 is equivalent to a quarter period oscillation (a $\pi/2$-pulse, in the notation of Ref.~\cite{doubleQD}). In step 1, this pulse creates an equal superposition state in the two dots.

We discuss now the second requirement (step 2). As we pointed out above, dot 0 is assumed to be a spin qubit, hence by hypothesis all the single qubit operations (including the sign flip) are already implemented in the original quantum computation scheme. This is one of the simplifying elements of our read-out scheme: we use a resource already present in the spin-qubit implementation. Hence we can view the present read-out scheme as a ``plug-in'' to be used in conjunction with an already existing spin-qubit implementation.

In the following we describe two possible setups which enact the local sign-flip operation required in step 2. The first is to use a magnetic tip AFM. The experiment described in Ref.~\cite{rugar}, of single spin detection using single magnetic resonance force microscopy (MRFM) opens up also the possibility of single spin manipulation using the same technique. In this case, a magnetic tip AFM can be applied locally to dot 0 (the spatial resolution achieved in Ref.~\cite{rugar} was 25 nm) in order to enact the sign-flip operation. The second setup employs the technique demonstrated by Kato {\it et al.} \cite{kato}, i.e., the coherent spin manipulation using electric fields in strained semiconductors. If the qubit dot (QD0) is defined in strained GaAs/InGaAs layers, the electron spin can be manipulated coherently using only applied electric fields.

\begin{figure}
\putfig{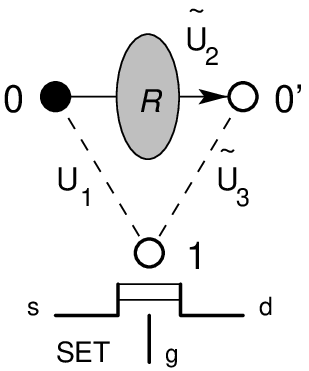}{2.5}
\caption{An alternative setup for measurement. Region $R$ acts on the electron spin as: $\ket{\up}\ra\, i\ket{\up}$, $\ket{\down}\ra -i\ket{\down}$. This can be produced in two ways by having either: (a) a Rashba active region, using top/bottom gates; or (b) a local static magnetic field. Since the electron tunnels completely from dot 0 to dot 0' (half Rabi oscillation), this introduces an extra $i$ factor on the upper branch.}
\label{rashba}
\end{figure}

A different way to solve the problem of fast switching of local fields required in step 2 is to use an alternative setup (see Figure \ref{rashba}). Instead of step 2 above (apply locally a spin-$\sigma_z$ on QD 0), the particle in dot 0 tunnels completely to a third QD, denoted by $0'$. Between QDs 0 and $0'$ there is a region $R$ enacting a spin rotation ${\sf R_z}(-\pi/2)= e^{-i\sigma_z\pi/2}= -i\sigma_z$. A complete tunneling between QDs 0 and 0' (half Rabi oscillation, $\ket{0}\ra\, i\ket{0'}$) introduces an extra $i$ factor in the orbital part, and hence the total transformation on the upper part of the interferometer will be: $\ket{\up; 0}\ra \ket{\up; 0'}$, $\ket{\down; 0}\ra -\ket{\down; 0'}$, which is effectively equivalent to $U_2$ in step 2 above.

The other steps remain the same: (i) a quarter of a Rabi oscillation ($\theta=\pi/4$) between QDs 0 and 1 ($U_1$ as above); (ii) complete tunneling between dots 0 and 0' through region $R$, enacting $\tilde{U}_2$; (iii) a quarter of a Rabi oscillation between 0' and 1 ($\tilde{U}_3$); (iv) charge measurement on dot 1 with a SET. Again, after applying $\tilde{U}_3 \tilde{U}_2 U_1$, a spin up (down) electron will be recovered with unit probability in QD 1 (0'), $\ket{\up; 0} \ra\ i\ket{\up; 1}$, $\ket{\down; 0} \ra -\ket{\down; 0'}$.

The region $R$ performing a spin rotation ${\sf R_z}(-\pi/2)$ can be implemented in two ways using either:\\
(a) a static magnetic field $B$ inducing a Zeeman splitting (e.g., produced by a micro-magnet situated in vicinity), or\\
(b) a static electric field $E$, if $R$ is a Rashba-active region (which should have a spin-orbit coupling controllable by top/bottom gates \cite{grundler, spin_orbit, aep_ri}). An estimation of the Rashba region length required for a $\pi/2$ spin rotation is $L= 58\,$nm in InAs ($\alpha= 4\times 10^{-11}$eVm \cite{grundler}) and $L= 250\,$nm in InGaAs/InAlAs ($\alpha= 0.93\times 10^{-11}$eVm \cite{Rashba1}).

By using static fields in region $R$ we eliminate the requirement of fast on/off switching of $U_2$ in step 2 and replace it with coherent control of tunneling $0 \ra 0'$ (which should be less restrictive from a technological point of view).

\begin{figure}
\putfig{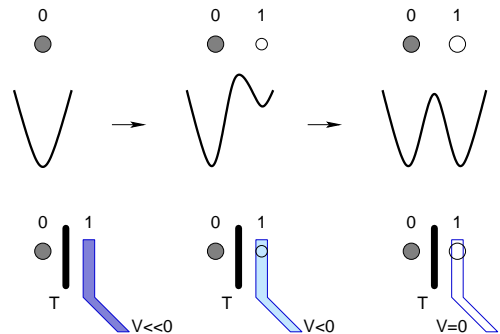}{6.5}
\caption{Creating a mode qubit by engineering the confining potential. The QDs are defined by top gates used to deplete the electrons in a 2DEG (not shown in the figure); a gate above (or near) dot 1 (blue polygon) is used to deplete the electrons in mode 1, physically destroying the second dot when it is not needed (by applying a negative potential $V$). The gate {\sf T} (black line) controls the tunneling between the two QDs.}
\label{well}
\end{figure}

Since charge coherence time is considerably shorter than spin coherence time, it is essential to have the ancilla dot 1 completely decoupled from the spin qubit (dot 0) during the whole quantum computation step. One way of achieving this is to ``create on demand'' dot 1, only when the measurement is required, e.g., by having a control gate close to QD 1. A large negative bias applied to this gate depletes the 2DEG electrons, physically ``destroying'' the dot (see Fig.~\ref{well}). Thus, prior to the measurement there is no orbital qubit but only the spin qubit used in computation.

\section{Further generalizations and conclusions}

The measurement method developed here can be extended to other configurations, e.g., optical lattices loaded with single atoms (in a Mott insulator phase) or Bose-Einstein condensates. The spin can be replaced by another internal degree of freedom elusive to direct measurement (denoted by $\Sigma$ in the following). The setup is similar: we can measure $\Sigma$ by mapping this internal degree of freedom into an external one (e.g., modes $k$) and subsequently measuring the modes. There are several assumptions behind this scheme: (i) the external degree of freedom $k$ is easier to measure than $\Sigma$; (ii) $\Sigma$ has only two possible values (or only two of them are relevant for the problem, e.g., the ground and first excited state of a particle in a potential well); and (iii) there exists an interaction which entangles $\Sigma$ and $k$.

The important point to notice here is that applying locally (i.e., only on mode 0) any unitary transformation which affects $\Sigma$ is equivalent to a controlled interaction between $k$ and $\Sigma$ ``qubits''. It is this interaction which maps the $\Sigma$-state into a $k$-state. Since $\Sigma$ and $k$ become entangled, measuring $k$ reveals the state of $\Sigma$.

Among the physical systems to which this architecture can be applied are atoms and Bose-Einstein condensates in optical lattices. The measurement step for (single) atoms in a optical lattice can be performed using fluorescence: an atom present (absent) in mode 1 will be seen as a bright (dark) spot under an appropriate laser illumination. The confinement potential shape (and hence the tunneling rate) can be controlled with counter-propagating laser beams \cite{charron}.

In conclusion, we have proposed a scheme for measuring the spin of a single electron confined to a QD, using the interplay of spin and orbital degrees of freedom and a subsequent charge measurement with a SET. Since spin-to-charge conversion is done with unit probability, single-shot measurement of a spin in a QD becomes in principle possible, using resources not more challenging than those available at present. Our design is theoretically and technologically compatible to state of the art schemes for universal QIP with electron spins in QDs. Other advantages include scalability and adaptability to purely electrical control. We believe that the present proposal is one of the simplest applications of the tunneling and spin rotation Hamiltonians to the task of single spin measurement.

\noindent {\bf Acknowledgments.}~We are grateful to P.~Zanardi and I.~D'Amico for useful comments and enlightening discussions. A.E.P.~thanks Gonville and Caius College, Cambridge for awarding her a Research Fellowship. R.I.~gratefully acknowledges funding by European Union project TOPQIP (contract IST-2001-39215).


\begin{thebibliography}{}
\bibitem{set} A SET is composed of a single electron box (white rectangle in Fig.\ref{spin_set}) situated between a source $s$ and a drain $d$. A plunger gate $g$ biases the single-electron island to work in the Coulomb blockade regime. The device is sensitive to a single electron charge in the nearby quantum dot. An alternative, more complex design, is to use two SETs (one for each QD) connected in anti-coincidence in order to avoid spurious (e.g., background charge) detections \cite{2SET_detector}.
\end{thebibliography}
\end{document}